\documentclass[11pt,a4paper,english,nofootinbib,,superscriptaddress]{revtex4}
\usepackage{lmodern}

\usepackage[T1]{fontenc}
\usepackage[latin9]{inputenc}
\usepackage{babel}
\usepackage{amsmath}
\usepackage{graphicx}
\usepackage{amssymb}
\usepackage{esint}
\usepackage[unicode=true, pdfusetitle,
 bookmarks=true,bookmarksnumbered=false,bookmarksopen=false,
 breaklinks=false,pdfborder={0 0 1},backref=false,colorlinks=false]
 {hyperref}
\usepackage{amsmath}
\usepackage{amssymb}
\def\b{\begin{equation}}
 \def\e{\end{equation}}

\makeatletter
\usepackage{latexsym}\usepackage{bm}

\makeatother

\begin{document}

\title{A Covariant Approach for Particle Creation in Non-flat Background}
\author{A. R. Ziyaee}
\affiliation{Department of Physics, Qom Branch, Islamic Azad University, Qom, Iran}

\author{M. Mohsenzadeh}
\email{Mohsenzadeh@qom-iau.ac.ir (Corresponding author)}
\affiliation{Department of Physics, Qom Branch, Islamic Azad University, Qom, Iran}
\author{E. Yusofi}
\email{e.yusofi@iauamol.ac.ir}
\affiliation{Department of Physics, Ayatollah Amoli Branch, Islamic Azad University, Amol, Iran}
\date{\today}

\begin{abstract}

\noindent \hspace{0.35cm} Krein approach is used to study the particle creation during quasi-de Sitter inflation in different background space-times. In the conventional method for calculating the created particles spectrum, the background space-time is automatically considered flat. Selecting a flat background poses two fundamental problems: First, the method of calculating is not covariant relative to the curved space-time. Second, the number of created particles becomes negative. Krein approach can be considered as a covariant method to calculate two-point functions in curved space-time. So we extend this method for particle creation during early cosmic inflation. As a new proposal, we choose the background vacuum based on the smallest number of created particles in that space-time. The calculations will show that in order to solve the above mentioned problems, the background space-time must be non-flat and the number of particles in the background must be minimum.
\end{abstract}

\maketitle

\section{Introduction and Motivation}
Even the simplest versions of the inflationary scenario appear to have the necessary mechanism to justify the homogeneity at large-scale universe and the tiny initial inhomogeneities to structure formation. So the spectrum of the inflationary models agree very closely with cosmological observations \cite{haw1, haw2, haw3, smo4}. Since, the primordial inflation has occurred in a dynamic and curved space-time, so the quantum mechanism governing this theory must be studied in a curved background \cite{bir5}.\\
The annals of the creation of gravitational particles in a curved space-time goes back to Schr\"odinger's early work in 1939 \cite{sch6}, and later was developed by Parker for an expanding background in the late 1960s \cite{par7, par8, par9}. A prominent trait of this method is using the mode functions that is compatible with the equation for the classical wave and also satisfies the Bogoliubov transformations. Despite many successful aspects of QFT in non-flat space-time as well as in the cosmic inflation scenario, the meaning of vacuum and particle in the dynamical non-flat space-time still has some fundamental problems. \cite{bir5, lin10, arm101}.\\
As we know, vacuum in flat space-time is quite distinct and unique, but in curved space-time, the definition of vacuum is not very obvious and there is anonymity in its selection. If for the very earliest times of the cosmic evolution, we consider a pure de Sitter inflation, for such a period under the de Sitter symmetry group there is a distinct class of invariant vacuum states. In fact, one of the essential foundations of the particle creation process in each era is how to choose and define the quantum vacuum state\cite{bir5}.\\
However, the accelerated expansion in the early universe was not occurred in a pure de Sitter space-time, because recent observational data motivate us to use quasi-de Sitter inflation instead of pure dS inflation\cite{bau11, pla12}. Consequently, it is more reasonable to describe physical early universe with a more physical initial states i.e. quasi-de Sitter vacuum. Therefore, we have introduced the co-called \emph{asymptotic- de Sitter} modes for the first time in \cite{moh13, moh131, yus14, yus15}, as the  main vacuum modes during inflation to study the subject of the particle creation. Compared to pure de Sitter inflation, particles created during quasi-de Sitter inflation have effective mass and non-minimal coupling with gravity that appears to be due to the change in the curvature of the space-time from de Sitter one \cite{nmc16, cjp17}. By using observational data, it has been shown that the background space-time for both early inflation and present accelerating universe is inferred to be quasi-de Sitter form dominated by a quasi-vacuum cosmic fluid \cite{mpl18}. \\
In some resent works, we used these alternative modes to calculate the higher order corrections to the spectra in a quasi-dS inflation \cite{cpc19}. Also, we have examined such asymptotic-de sitter modes in the light of the planck data \cite{cpc20}. In the present work, inspired by the Krein approach in $QFT$ \cite{tak21, tak22, tak23} and using of the curved background for gravitational waves instead to flat background \cite{car24}, we will obtain a covariant relation to calculate of spectrum of created particles. So, in Sec. 2 we reintroduce the asymptotic de Sitter vacuum modes during inflation. In the main Sec. 3, we introduce Krein approach in the cosmological spectra calculations. By using asymptotic-de Sitter modes as the fundamental initial states during inflation, we obtain the spectra of created particles in different space-time with Krein and standard approach . The comparison between the two approaches is performed by drawing some relevant diagrams. In the final section conclusions and outlooks are presented.\\

\section{Asymptotic Vacuum Modes During Inflation}
The recent cosmological observations motivate us to use perturbed form of diagonal metric in Newtonian gauge as follows \cite{bau11},
\begin{align} \label{eq:line-element}
ds^{2}= a^{2}(\tau)(-(1+2\Phi){d\tau}^2+(1-2\Psi){d\textbf{x}}^2),
\end{align}
where $\Phi$ and $\Psi$ are the gauge-invariant potentials.  Due to Einstein equation, the perturbed parts of metric and the inflaton field are related to in gauge invariant formalism by using Mukhanov
variables $v$ and $z$. The dynamics of quantum inflaton field fluctuations are governed by following action \cite{asl25}
\begin{equation} \label{eq:action}
S=\frac{1}{2}\int d^3xd\tau\left((v')^2-(\nabla
v)^2+\frac{z''}{z}v^2\right).
\end{equation}
Therefore, the equation of motion in Fourier space for primordial scalar
perturbations is\cite{bau11}
\begin{equation}\label{muk7}
{v''}_{k}+\left(k^{2}-\frac{z^{\prime \prime}}{z}\right)v_{k}=0,
\end{equation}
where derivatives are respect to conformal time
$\tau$. Also $v_{k}$ is the Fourier mode of quantum field.
\\
We have $\frac{z^{\prime \prime}}{z}= \frac{2c}{\tau^2} $, for the dynamical background during inflation, in the equation (\ref{muk7}),  where $c$ is given by \cite{yus14, asl25},
 \begin{equation}
 \label{alf26}  c=\frac{4\nu^2-{1}}{8}.
 \end{equation}
 The general solutions of (\ref{muk7}), can be written as the first and second kind of the Hankel functions $ H_{\nu}^{(1)} $ and $ H_{\nu}^{(2)}$, respectively \cite{yus14, yus15}:
\begin{equation}
\label{Han22}
\upsilon_{k}=\frac{\sqrt{\pi \eta}}{2}\Big(A_{k}H_{\nu}^{(1)}(|k\eta|)+B_{k}H_{\nu}^{(2)}(|k\eta|)\Big). \end{equation}
 For the first time in the \cite{yus14}, we have been used the asymptotic expansion of the Hankel functions at the early time limit $k\tau\gg 1$, in term of $c$ as follows
 \begin{equation} \label{gen27}  \upsilon^{gen}_{k} = A_{k}\frac{e^{-{i}k\tau}}{\sqrt{k}}\big(1-i\frac{c}{k\tau}-\frac{d}{k^2\tau^2}-...\big)+
 B_{k}\frac{e^{{i}k\tau}}{\sqrt{k}}\big(1+i\frac{c}{k\tau}-\frac{d}{k^2\tau^2}+...\big),
 \end{equation}
where $d={c(c-1)}/{2}$. The positive frequency solution as asymptotic-dS vacuum is given by
\begin{equation}
\label{mod28}  \upsilon_{k}^{adS}=\frac{e^{-{i}k\tau}}{\sqrt{k}}\left(1-i\frac{c}{k\tau}-\frac{d}{k^2\tau^2}-...\right).
 \end{equation}
If we consider the special case($\nu={3}/{2}$ or $c = 1$), the general form of the mode functions (\ref{mod28}) reduces to first order mode i.e. the pure dS mode \cite{bun26}:
  \begin{equation}
 \label{Bun29} \upsilon_{k}^{dS}=\frac{1}{\sqrt{k}}(1-\frac{i}{k\tau})e^{-ik\tau}.
\end{equation}
For very early universe with pure de Sitter inflation we have $a(t)= e^{Ht}$ or $ a(\tau)=-\frac{1}{{H}\tau}$, with $H=constant$. The best values of $\nu$ which are consisted with the latest observational data \cite{pla12}, are $1.513\leq \nu \leq 1.519$ \cite{nmc16}. This range of $\nu$ motivated us to the departure from dS inflation with pure dS mode $(c = 1)$ to the quasi-dS inflation with asymptotic-dS modes $(c \neq 1)$. Therefore we can use asymptotic-dS modes (\ref{mod28}) instead pure dS mode (\ref{Bun29}) as the main physical mode for the early inflation that is consist with observational data.

\section{Krein Approach in Cosmological Calculations}
The complete solution in the Krein approach is made up of a combination of two positive and negative norm states\cite{tak21, moh29, moh30},
 \b \phi(x)=\frac{1}{2}
[\phi_p(x)+\phi_n(x)],\e where
$$ \phi_p(x)=(2\pi)^{-3/2}\int d^{3}k [p(k)u_{k,p}(t)e^{i\textbf{k.x}}+p^{\dag}(
k)u_{k,p}^*(t)e^{-i\textbf{k.x}}],$$ $$
\phi_n(x)=(2\pi)^{-3/2}\int d^{3}k [n(
k)u_{k,n}(t)e^{-i\textbf{k.x}}+n^{\dag}(k)u_{k,n}^*(t)e^{i\textbf{k.x}}],$$
$ p(k) $ and $ n(k) $ are two independent operators in spaces with positive and negative norms, respectively. The operators obey the commutation rules as follows \b [p(k),p^{\dag}(k')]=\delta(k-k') ,\;\; [n(
k),n^{\dag}( k')]=-\delta( k-k') .\e All of other commutation
relations equal to zero. The vacuum state $\mid \Omega>$
is then defined by \b p^{\dag}(k)\mid \Omega>= \mid 1_{
k}>;\;\;p(k)\mid \Omega>=0, \e \b n^{\dag}( k)\mid \Omega>= \mid
\bar1_{k}>;\;\;n( k)\mid \Omega>=0, \e where $\mid 1_{k}> $ is
a one particle (physical) state and $\mid \bar1_{k}>$ is a one
(un-physical) state. For more information on Krein method and some of its results, see references
\cite{tak21, tak22, tak23, moh29, moh30}.\\
In Hilbert's method, the expectation value for the energy-momentum tensor become infinite. The normal ordering method is used to obtain a finite answer. But in curved space-time this method is not suitable and the following remedy is usually used for a finite solution\cite{bir5},
\begin{equation}
\label{equ11} \langle\Omega|:T_{\mu\nu}:|\Omega\rangle=\langle\Omega|T_{\mu\nu}|\Omega\rangle-\langle0|T_{\mu\nu}|0\rangle,
 \end{equation}
In the second sentence of the equation (\ ref {Equ11}), which relates to the flat background, the negative sign can be interpreted as the negative norm effect in Krein method. So we can have the following equivalence. \cite{tak22,moh29}:
\begin{equation}
\label{equ12}
\langle\Omega|{:\phi^{2}:}|\Omega\rangle_{Krein}= \langle
\phi^{2}\rangle_p+\langle \phi^{2}\rangle_n = \langle
\phi^{2}\rangle_{phy} -\langle \phi^{2}\rangle_{bac},
\end{equation}

Inspired by (\ref{equ11}) and (\ref{equ12}), the following definition for the renormalized spectra of created particlescan be defined as follows,
\begin{equation}
 \label{equ142}
 {\langle N\rangle}_{Krein}= {\langle N\rangle}_{p}+ {\langle N\rangle}_{n}= \langle\Omega_{phy}|N|\Omega_{phy}\rangle-\langle\Omega_{bac}|N|\Omega_{bac}\rangle,
\end{equation}
where $|\Omega_{phy}\rangle$ and $|\Omega_{bac}\rangle$ are the positive ($\equiv$ physical) vacuum state and negative ($\equiv$ bakground)vacuum state, respectively. Since, the early universe is expanded with a quasi-de Sitter inflation \cite{bau11}, so we choose asymptotic-dS modes(\ref{mod28}) as the initial physical vacuum, but for the background (un-physical) vacuum we have both flat and non-flat options. In the next subsections, we will use (\ref{gen2}) and calculate the number of created particles in some specific space-times by the standard and Krein approach.
\section{Particle Creation during Asymptotic Inflation}
Using the Bogoliubov coefficients, the number of created particles for the vacuum mode $ \upsilon_{k} $ is calculated as follows \cite{mij27, per28},
 \begin{equation} \label{equ13} \langle N \rangle= -\frac{1}{2}+\frac{1}{4\omega_{k}(\eta)}|{\upsilon}'_{k}(\eta)|^{2}+\frac{\omega_{k}(\eta)}{4}|{\upsilon}_{k}(\eta)|^{2}. \end{equation}
\begin{figure}[h]
\centering
\includegraphics[width=3in]{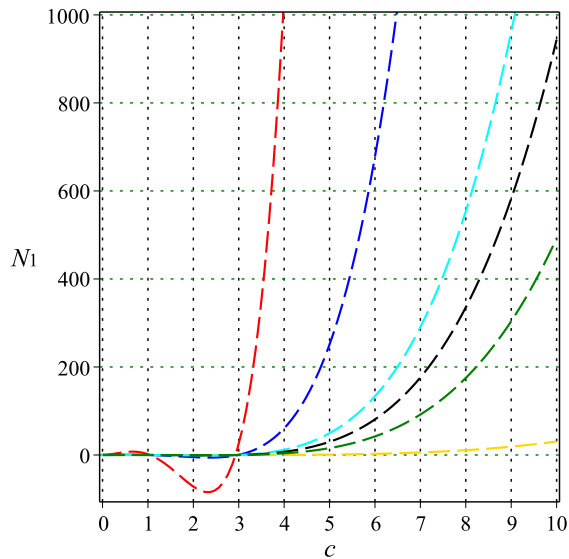}
\caption{The number of created particles ${\langle N_1 \rangle}_{K} = \langle{N}\rangle_{adS}$ in term of parameter $c$ in \emph{standard} approach with $flat$ background is plotted for different assymptotic-dS space-times for $k\tau=-5$ (red), $k\tau=-10$ (blue), $k\tau=-15$ (gray), $k\tau=-17$ (black), $k\tau=-20$ (green), and for $k\tau=-40$ (gold).}
\label{Fig1}
\end{figure}
It is generally expected that particles will create due to changes in the gravitational field in the expanding background of the universe\cite{par7}. In fact, how to select a specific vacuum plays a key role in the number of created particles\cite{bir5}. Our general mode (\ref{mod28}) is not only time dependent but also dependent on the parameter $c$. Therefore, any selection of $c$ yields a different set of vacuum modes $\upsilon_{k}$. Anyway, by using equation (\ref{equ13}), the number of created particles for the vacuum modes (\ref{mod28})has been calculated as the follows \cite{nmc16},

$$ \langle N \rangle_{adS} = -\frac{1}{2}+\frac{1}{4}|k^{2}\tau^{2}-2c|^{1/2}\big[\frac{1}{k\tau}+\frac{c}{k^{3}\tau^{3}}
+\frac{d^{2}}{k^{5}\tau^{5}}+...\big]$$
\b \label{gen2}
+\frac{1}{4|k^{2}\tau^{2}-2c|^{1/2}}\big[k\tau-\frac{c}{k\tau}+\frac{(c-d)^{2}}{k^{3}\tau^{3}}
+\frac{4d^{2}}{k^{5}\tau^{5}}+...\big].
\e
\subsection{Calculations for Special Cases}
\subsubsection{Flat space-time with $c=0$}
The number of created particles for flat vacuum mode,
\begin{equation} \label{mod281}
\upsilon^{flat}_{k}=\frac{e^{-{i}k\tau}}{\sqrt{k}},
 \end{equation}
obtained as follow,
\b \langle{N}\rangle_{flat}=0.
\e
This is an obvious result to the conventional method because background space-time is chosen flat in the standard particle theories.
\subsubsection{de Sitter space-time with $c=1$}
For pure dS vacuum mode (\ref{Bun29}), the result is as follows,
\b \langle{N}\rangle_{dS} = -\frac{1}{2}+\frac{1}{4}|k^{2}\eta^{2}-2|^{1/2}(\frac{1}{k\eta}+\frac{1}{k^{3}\eta^{3}})+\frac{1}{4|k^{2}\eta^{2}
-2|^{1/2}}(k\eta-\frac{1}{k\eta}+\frac{1}{k^{3}\eta^{3}}).
\e
 This result previously has been obtained in \cite{per28}.\\
\subsubsection{Quasi-de Sitter space-time with $ 1 < c < 3$}
In order to illustrate the phenomenon of particle creation in dynamical inflationary background , let us consider the quasi-dS space-time, that observationally is very significant. In that case we choice $c = 2.34$, with $d = 1.5678$,
\begin{equation} \label{mod291}  \upsilon_{k}^{qdS}=\frac{e^{-{i}k\tau}}{\sqrt{k}}\left(1-i\frac{2.34}{k\tau}-\frac{1.5678}{k^2\tau^2}\right),
 \end{equation}
and the $minimum$ number of the created particles up to second order $\frac{1}{k\eta}$ of mode (\ref{mod291}) is
$$ \langle{N}\rangle_{qdS} = \langle{N}\rangle_{min} = -\frac{1}{2}+\frac{1}{4}|k^{2}\tau^{2}-4.68|^{1/2}\big[\frac{1}{k\tau}+\frac{2.34}{k^{3}\tau^{3}}
+\frac{2.457996840}{k^{5}\tau^{5}}\big]$$
\b \label{gen5}
+\frac{1}{4|k^{2}\tau^{2}-4.68|^{1/2}}\big[k\tau-\frac{2.34}{k\tau}+ {\frac {0.5962928400}{{k}^{3}{\tau}^{3}}}+\frac{9.831987360}{k^{5}\tau^{5}}\big].
\e
Note that the correction terms in (\ref{gen5}), were very small during early time, but grow at later time.

\section{Gravitational Particles Creation with Krein Approach}
\subsection{Creation on the flat background}
In the gravitational wave point of view, we have,
\begin{equation}
 \label{equ15-1}
 h^{i}_{\mu\nu} = g^{phy}_{\mu\nu} - g^{bac}_{\mu\nu},
\end{equation}
where, $h^{i}_{\mu\nu} = \delta g_{\mu\nu}$ is the metric perturbations. Accordingly for the flat background, we can write,
\begin{equation}
 \label{equ151}
 h^1_{\mu\nu} = g^{adS}_{\mu\nu} - g^{flat}_{\mu\nu},
\end{equation}
and the number of created gravitational particles is given by,
\begin{equation}
 \label{equ152}
 {\langle N_1 \rangle}_{K} = \langle{N}\rangle_{adS}-\langle{N}\rangle_{flat}.
\end{equation}
But the relation $(\ref{equ152})$ has two major problems, first of all, it is not in covariant form. Second, the number of created particles in the interval $(1 < c < 3)$ is negative, and this case is completely unacceptable (see figure 1).
\begin{figure}[h]
\centering
\includegraphics[width=3in]{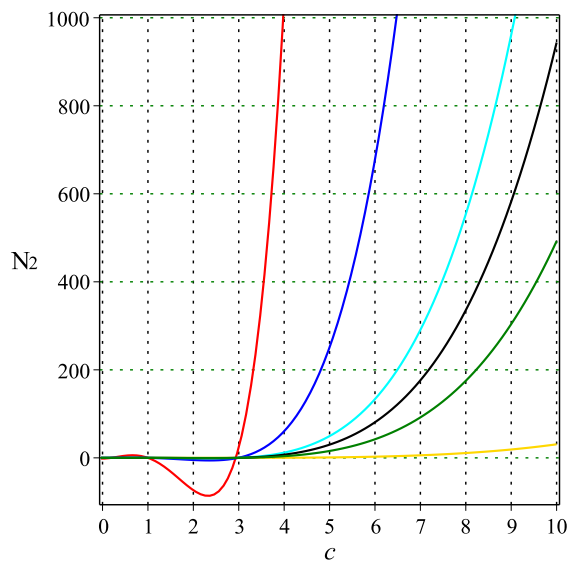}
\caption{The number of created particles ${\langle N_2 \rangle}_{K} = \langle{N}\rangle_{adS}- \langle{N}\rangle_{dS}$ in term of parameter $c$ in \emph{Krein} approach with $dS$ background is plotted for different assymptotic-dS space-times for $k\tau=-5$ (red), $k\tau=-10$ (blue), $k\tau=-15$ (gray), $k\tau=-17$ (black), $k\tau=-20$ (green), and for $k\tau=-40$ (gold).}
\label{Fig2}
\end{figure}
\begin{figure}[h]
\centering
\includegraphics[width=3in]{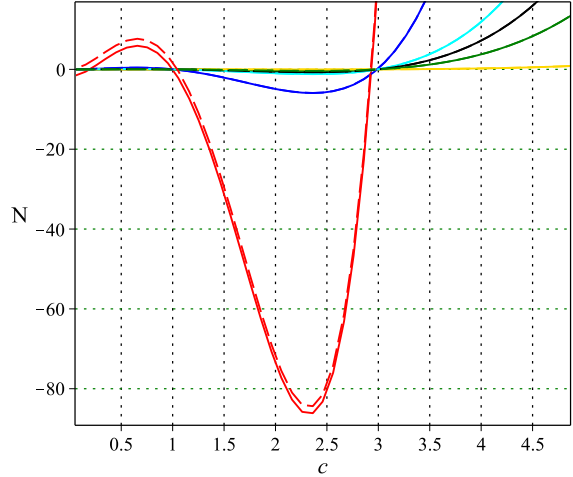}
\caption{Comparison the number of created particles $N_1$ (dash line) and $N_2$ (solid line) in a single plot.}
\label{Fig3}
\end{figure}
\subsection{Creation on the dS background}
But, in the novel point of view in the background field method we consider $g^{cur}_{\mu\nu} \equiv g^{bac}_{\mu\nu}$ that is consist with both of quasi-de Sitter inflation and first order cosmological perturbations. The metric perturbations for the quasi-de Sitter inflation on the dS background
\begin{equation}
 \label{equ153}
h^2_{\mu\nu}= g^{adS}_{\mu\nu}- g^{dS}_{\mu\nu},
\end{equation}
and the number of created gravitational particles is given by,
\begin{equation}
 \label{equ154}
 {\langle N_2 \rangle}_{K}=\langle{N}\rangle_{adS}-\langle{N}\rangle_{dS},
\end{equation}
Note that when the background metric is flat, the results of Krein method  is the same as 'Hilbert' (standard) method. Note that the relation $(\ref{equ154})$ is in covariant form, but still the number of created particles in the interval $(1 < c < 3)$ is negative.\\
\begin{figure}[h]
\centering
\includegraphics[width=3in]{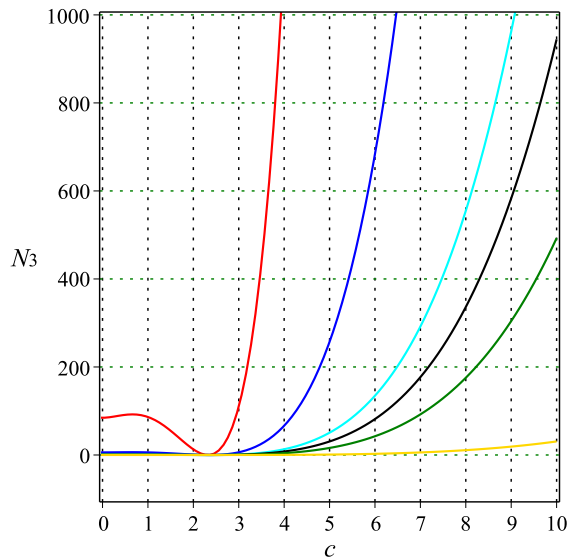}
\caption{The number of created particles ${\langle N_3 \rangle}_{K} = \langle{N}\rangle_{adS}- \langle{N}\rangle_{min}$ in term of parameter $c$ in \emph{Krein} approach with $qdS$ bacground is plotted for different assymptotic-dS space-times for $k\tau=-5$ (red), $k\tau=-10$ (blue), $k\tau=-15$ (gray), $k\tau=-17$ (black), $k\tau=-20$ (green), and for $k\tau=-40$ (gold).}
\label{Fig4}
\end{figure}
\subsection{Creation on a background with minimum number of particles}
As one can see from figure 1, 2 and 3, both of the relations $(\ref{equ152})$ and $(\ref{equ154})$  can not solve the problem of negative particle number. But the reason for the negative number of particles can be found in Figures 1, 2 and 3. In these figures, for a special value of parameter $c$ i.e. $c = 2.34$, the number of created particles has the lowest value. Interestingly, the number of particles in this case is also smaller than the number for flat and de Sitter space-times, i.e.  $c = 0$ and $c = 1$, respectively. Therefore, it seems that the problem of negative number of created particles in the proposed method can be eliminate, if we consider $\upsilon_{k}^{qdS}$ as a main physical background. For this purpose we consider $g^{qdS}_{\mu\nu} \equiv g^{bac}_{\mu\nu}$. Therefore, we can write,
\begin{equation}
 \label{equ155}
 h^3_{\mu\nu}= g^{adS}_{\mu\nu}- g^{qdS}_{\mu\nu},
\end{equation}
and the number of created gravitational particles is given by,
\begin{equation}
 \label{equ156}
 {\langle N_3 \rangle}_{K}=\langle{N}\rangle_{adS}-\langle{N}\rangle_{min},
\end{equation}
In this case, the result of the Krein method is no longer the same as the two previous cases, and the choice of the quasi-de Sitter background changes the final result and removes of negative number problem for created particles as is show in figure 4. Also, the final result is new and is consistent with the first order cosmological perturbations as well as observational quasi-dS inflation.\\
By comparing the 3 methods $A.$, $B.$ and $C.$, it can be concluded that in addition to the main physical vacuum mode, the choice of background modes plays a key role in the calculation of the number of created particles. So in addition to the concept of particle, the creation of particles in the curved space-time is dependent on the choice of the vacuum, and with the change of background space-time during inflation, the results will change (Please compare figures 1, 2 and 4). Also, according to both of figures 1, 2 and 4, the effect of high-order terms and the parameter $c$ are very small at the initial time ($|k\tau| \gg 0 $), but these effects become more pronounced over time ($|k\tau| \rightarrow 0$).\\
\\
\section{Conclusions}
 A form of covariant approach in curved space-time has been used to study the particle creation during quasi-de Sitter inflation in different asymptotic-de Sitter background space-times. Krein approach has been considered as a covariant method for calculation of two-point functions for quantum fields in curved space-time. So we extend this method in the essue of early universe cosmology to calculate the spectrum of created particles during early inflation. Calculations have been shown that the effect of our proposed method appears only when both sets of physical and background vacuum modes are chosen from different non-flat space-times.\\
 Addition to the standard method to calculate of power spectrum, we have proposed that the initial vacuum modes are non-flat at early time, that was asymptotically de Sitter at very early time limit. The calculations performed in quasi-dS space-time showed that the minimum of created particles where not related to the flat space-time, but in the range of $(1 < c < 3)$, the number of created particles by asymptotic-dS vacuum mode are less than flat one. Especially in particular case $c = 2.34$, we have the minimum of the created particles number. As a new proposal, we have chosen the background vacuum based on the smallest number of created particles in that space-time. Therefore, we have selected case $c = 2.34$, as a background space-time for vacuum state instead of the flat case i.e. $c = 0$, and with this trick we were able to solve the problem of negative number of created particles during quasi-dS inflation.\\
 \\
\noindent{\bf{Acknowlegements}}: This work has been supported by the Islamic Azad University, Qom Branch, Qom, Iran.

\end{document}